\documentclass[aps,prb,floatfix,twocolumn,showpacs,superscriptaddress]{revtex4-2}
\usepackage{graphicx}
\usepackage{amsmath,amssymb}
\usepackage{flushend}
\usepackage[utf8]{inputenc}
\usepackage[colorlinks = true,
linkcolor = blue,
urlcolor  = blue,
citecolor = blue,
anchorcolor = blue]{hyperref}
\usepackage{array}
\usepackage{bbold}
\usepackage{float}
\usepackage{enumitem}
\usepackage{braket}
\usepackage[mathscr]{euscript}
\usepackage{booktabs}
\usepackage{appendix}
\mathchardef\Re="023C
\mathchardef\Im="023D
\usepackage{color}
\usepackage{times}

\begin{document}
\title{Circuit realisation of a two-orbital non-Hermitian tight-binding chain}
\author{Dipendu Halder}
\email[]{h.dipendu@iitg.ac.in}
\affiliation{Department of Physics, Indian Institute of Technology Guwahati-Guwahati, 781039 Assam, India}
\author{Ronny Thomale}
\email[]{rthomale@physik.uni-wuerzburg.de}
\affiliation{Institute for Theoretical Physics and Astrophysics, University of Würzburg, D-97074 Würzburg, Germany}
\affiliation{Department of Physics and Quantum Centers in Diamond and Emerging Materials (QuCenDiEM) Group, Indian Institute of Technology Madras, Chennai, India}
\author{Saurabh Basu}
\email[]{saurabh@iitg.ac.in}
\affiliation{Department of Physics, Indian Institute of Technology Guwahati-Guwahati, 781039 Assam, India}

\begin{abstract}
We examine a non-Hermitian (NH) tight-binding system comprising of two orbitals per unit cell and their electrical circuit analogues. We distinguish the $\mathcal{PT}$-symmetric and non-$\mathcal{PT}$-symmetric cases characterised by non-reciprocal nearest neighbour couplings and onsite gain/loss terms, respectively. The localisation of the edge modes or the emergence of the topological properties are determined via the maximum inverse participation ratio, which has distinct dependencies on the parameters that define the Hamiltonian. None of the above scenarios exhibits the non-Hermitian skin effect. We investigate the boundary modes corresponding to the topological phases in a suitably designed electrical circuit by analyzing the two-port impedance and retrieve the admittance band structure of the circuit via imposing periodic boundary conditions. The obtained results are benchmarked against the Hermitian version of the two-orbital model to compare and differentiate from those obtained for the NH variants.
\end{abstract}

	\flushbottom
	\maketitle
	
	\thispagestyle{empty}
	
	\section{Introduction}
	
The mathematical discipline of topology, dedicated to the investigation of geometric properties preserved under continuous transformations, has witnessed a remarkable convergence with the field of condensed matter physics in the past few decades.
The seminal proposal of topological states in polyacetylene chain \cite{PhysRevLett.42.1698} and the presence of quantised edge modes in the quantum Hall effect \cite{PhysRevLett.45.494} marked pivotal turning discoveries, highlighting the role of topology in characterising novel phases of matter.
Thus, a prominent subset of materials, known as `topological insulators', has emerged as a central focus in contemporary research due to their unique electronic properties, particularly the presence of gapless edge modes that persist even under alteration of the band properties as long as the spectral gap stays open \cite{RevModPhys.82.3045, RevModPhys.83.1057, RevModPhys.88.021004}.
At the core of this inquiry lies the classification of distinct phases based on the underlying symmetries of the system, such as time-reversal symmetry (TRS), particle-hole symmetry (PHS), and chiral symmetry (CS).
The culmination of these efforts has led to the establishment of a comprehensive classification scheme for the topological insulators, commonly referred to as the `ten-fold' way \cite{RevModPhys.88.035005, Ryu_2010}.
	
Although we are trained to think about Hermitian systems with the energy being observable, and always assuming real values, there is significant enthusiasm for non-Hermitian (NH) systems as well.
Building upon foundational work by Bender and Boettcher \cite{PhysRevLett.80.5243}, it has been demonstrated that NH systems, exhibiting parity and time-reversal ($\mathcal{PT}$) symmetry may manifest real eigenspectra, despite the shift from the Hermitian paradigm.
The parity operator ($\mathcal{P}$) acts on the spatial part of a given state, resulting in an inversion of the position,
    \begin{equation*}
        \mathcal{P}:(x,y,z,t)\rightarrow(-x,-y,-z,t),
    \end{equation*}whereas the time-reversal operator ($\mathcal{T}$) reverses the temporal part, resulting in,
    \begin{equation*}
        \mathcal{T}:(x,y,z,t)\rightarrow(x,y,z,-t).
    \end{equation*}
Thus, the $\mathcal{PT}$ operator as a whole does not flip the momentum of the particle, as each of them causes a sign change of the momentum.

In particular, the interplay of the topology and NH systems \cite{PhysRevLett.116.133903, PhysRevLett.120.146402} has emerged as a captivating frontier in the realm of condensed matter physics.
Particularly, the `ten-fold way' is upgraded to a $38$-fold classification scheme \cite{PhysRevX.9.041015}.
This fortuitous interplay has unveiled an array of fascinating physical phenomena, notably including the non-Hermitian skin effect (NHSE), wherein the bulk eigenstates predominantly localise near the system boundaries and display pronounced sensitivity to boundary conditions \cite{PhysRevLett.121.086803, PhysRevLett.121.026808, PhysRevB.97.121401, PhysRevB.99.201103, PhysRevLett.124.056802, YUCE2020126094, PhysRevLett.124.086801, PhysRevB.102.205118}.
Consequently, the conventional concept of the Brillouin zone is challenged by the non-Bloch theory \cite{PhysRevLett.123.066404}, introducing a generalised Brillouin zone approach.
Exceptional points \cite{PhysRevB.97.121401, Heiss, PhysRevLett.118.040401}, where the eigenvalues and eigenfunctions of the `defective' Hamiltonian coalesce, have emerged as fundamental features in studying NH systems.
Thus, NH systems present a vast platform to explore the connection between topology and non-Hermiticity.
 
The experimental exploration of the topological systems has been materialised in diverse physical settings, spanning ultra-cold atoms in optical lattices \cite{El-Ganainy, Eichelkraut}, electronic \cite{PhysRevLett.114.173902, PhysRevX.5.021031, Imhof}, mechanical \cite{Wang}, and acoustic \cite{Fleury2015, PhysRevApplied.16.057001} systems.
Remarkably, electrical circuits have surfaced as a versatile platform for investigating the topological properties, owing to the inherent simplicity and adaptability in circuit design \cite{PhysRevB.99.161114, PhysRevB.100.045407, Lee, PhysRevResearch.2.023265, Helbig, PhysRevB.103.014302, PhysRevResearch.3.023056, PhysRevResearch.4.043108, PhysRevB.107.085426, Ganguly}.
Comprising of essential electronic components, these circuits produce results that solely depend on the connectivity and periodicity of the circuit elements, thereby mirroring the behavior of tight-binding (TB) systems.
As a result, these readily accessible and cost-effective electrical circuits provide a convenient and effective technique for experimental verification of quantum systems.
	
Our investigation centres around a comprehensive exploration of the intricate relationship between topology and NH systems through a one-dimensional TB system, comprising of two orbitals, $A$ and $B$, per unit cell, and their implementation in electronic circuits.
Additionally, we explore NH variants of the traditional (Hermitian) system, characterised by the presence (or absence) of $\mathcal{PT}$ symmetry.
The NH models are introduced through onsite gain/loss terms and non-reciprocal coupling strengths, leading to a detailed examination of their localisation and spectral behaviour, respectively, under both open boundary condition (OBC) and periodic boundary condition (PBC).
Our paper is organised as follows.
Section \ref{sec2} presents our results in a systematic sequence.
To summarise, we explore the properties of three TB models.
We begin with an in-depth discussion of the Hermitian case, followed by thoroughly exploring the non-$\mathcal{PT}$-symmetric and $\mathcal{PT}$-symmetric cases.
This sequential approach provides a clear understanding of the distinct behaviours exhibited by each of the above scenarios.
Subsequently, we delve into a comprehensive topolectrical analysis of each of the TB models.
This involves the construction of electronic circuits and simulating their observational properties, like impedance profile (IP) and admittance band structure (ABS), providing a bridge between theoretical models and experimental setups.
In section \ref{sec3}, we summarise the results.
	
	\section{\label{sec2}Models and Results}
	
	\subsection{Hermitian circuit}
	
We begin by considering a straightforward yet inclusive Hermitian ladder-type lattice model, where each unit cell accommodates an atom comprising of two different orbitals, that are denoted as $A$ and $B$ orbitals.
The corresponding Hamiltonian is expressed as follows:
	\begin{align}
		H_1=\sum_{i=1}^{L}&\bigg[\epsilon(\hat{a}_{i}^{\dagger}\hat{a}_{i}-\hat{b}_{i}^{\dagger}\hat{b}_{i})-t_{AB}\hat{a}_{i}^{\dagger}\hat{b}_{i-1}\bigg]+\nonumber\\&\sum_{i=1}^{L-1}\bigg[-t(\hat{a}_{i}^{\dagger}\hat{a}_{i+1}-\hat{b}_{i}^{\dagger}\hat{b}_{i+1})+t_{AB}\hat{a}_{i}^{\dagger}\hat{b}_{i+1}\bigg]+\textrm{H.c.}
		\label{eq:Ham1}
	\end{align}
Here, $\pm\epsilon$ symbolises the onsite potentials pertaining to the $A$ and $B$ orbitals, respectively. The hopping strengths are denoted by $-t$ ($t$) and $t_{AB}$ ($-t_{AB}$), governing the inter-cellular transitions $A^{i-1} \leftrightarrow A^i$ ($B^{i-1} \leftrightarrow B^i$) and $A^{i-1} \leftrightarrow B^i$ ($B^{i-1} \leftrightarrow A^i$), correspondingly.
$A^i$ and $B^i$ refer to the $A$ and $B$ orbitals at the $i^{th}$ unit cell, and $L$ denotes the system size or equivalently, the total count of unit cells.
The operators $\hat{a}_{i}$ ($\hat{a}_{i}^{\dagger}$) and $\hat{b}_{i}$ ($\hat{b}_{i}^{\dagger}$) denote the annihilation (creation) operators for spinless fermions that pertain to the $A$ and $B$ orbitals, respectively, at the $i^{th}$ unit cell.
The intra-orbital hopping term (between $A$ and $B$ orbitals within the same unit cell) is ignored.
In the context of PBC, the Hamiltonian in Eq.\eqref{eq:Ham1} can be stated in the ensuing Bloch form as,
	\begin{equation}
		h_1(k)=\begin{pmatrix}
			\epsilon-2t\cos k & 2it_{AB}\sin k\\-2it_{AB}\sin k & -\epsilon+2t\cos k
		\end{pmatrix}.
		\label{eq:kspace_1}
	\end{equation}
In the present case, the system has TRS, which is given by the following relation for any Bloch Hamiltonian, $h(k)$ \cite{Ryu_2010},
	\begin{equation}
		\mathcal{T}h(k)\mathcal{T}^{-1}=h(-k);\quad\mathrm{with}\quad\mathcal{T}^2=\pm1,
		\label{eq:TRS}
	\end{equation}
where $\mathcal{T}=U_T\mathcal{K}$, with $U_T$ and $\mathcal{K}$ being a unitary and the complex conjugation operator, respectively, which makes $\mathcal{T}$ anti-unitary in nature.
For systems consisting of spinless fermions (present case), $\mathcal{T}$ is nothing but the complex conjugation operator $\mathcal{K}$ and $\mathcal{T}^2=1$.
$h_1(k)$ satisfies the Eq.\eqref{eq:TRS} and hence obeys TRS.
	
In the same way, the system also possesses PHS, which is written as,
	\begin{equation*}
		\mathcal{C}h(k)\mathcal{C}^{-1}=-h(-k);\quad\mathrm{with}\quad\mathcal{C}^2=\pm1.
		\label{eq:PHS}
	\end{equation*}
The anti-unitary PHS operator $\mathcal{C}=U_C\mathcal{K}$ anti-commutes with the Bloch Hamiltonian $h_1(k)$, where $U_C=\sigma_x$ for the present case.
Evidently, the chiral symmetry (CS) is also there since the CS operator ($\Gamma$) is nothing but $\Gamma=\mathcal{T}\mathcal{C}$.
These properties allow us to conclude that the system falls in the class $\pmb{\mathrm{BDI}}$ in $\pmb{\mathrm{AZ}}$ symmetry classification \cite{RevModPhys.88.035005}.
	\begin{figure}[h!]
		\centering
		\includegraphics[height=8cm, width=9cm]{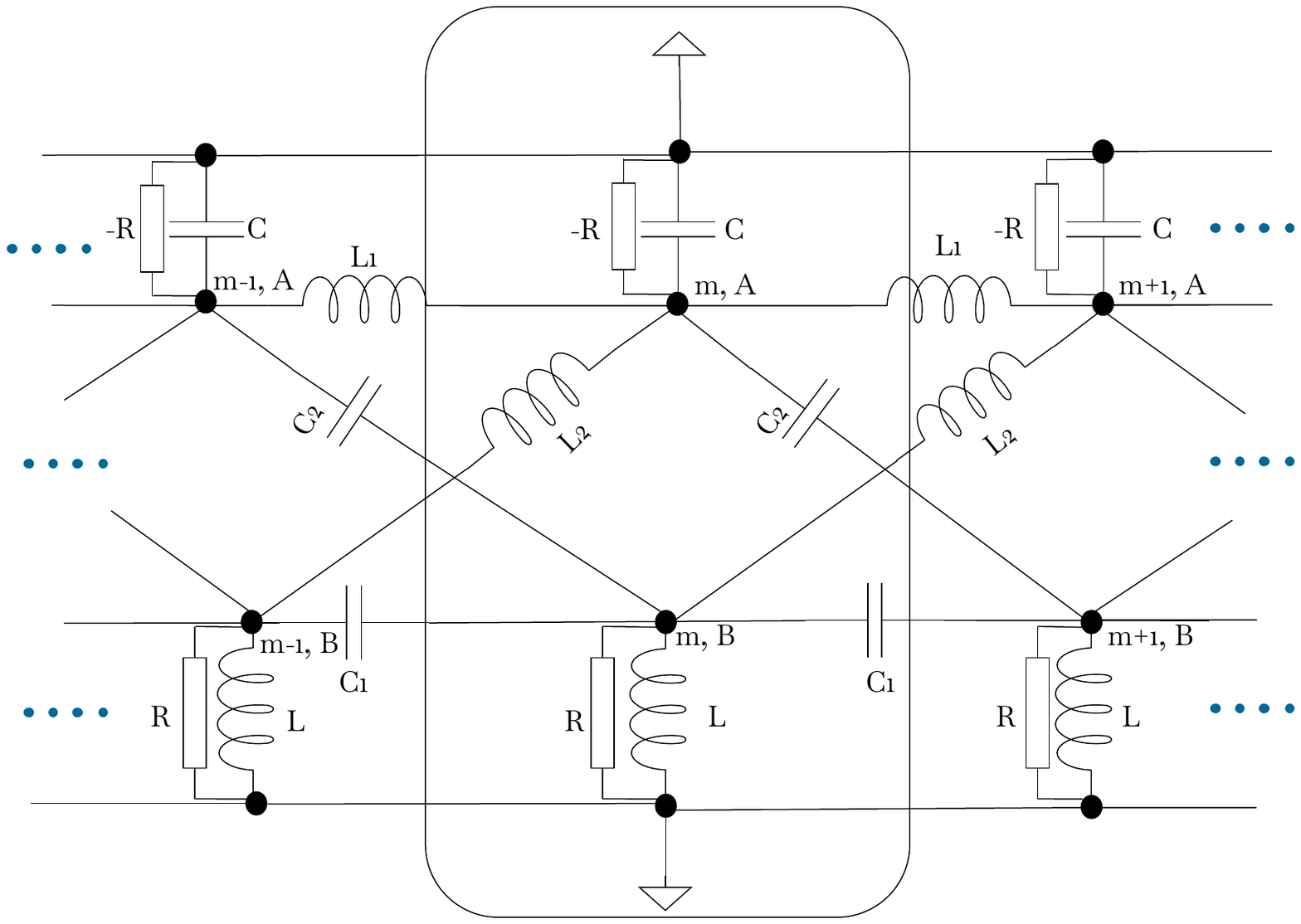}
		\caption{Circuit diagram for $3$ subsequent unit cells corresponding to the Hermitian model in Eq.\eqref{eq:Ham1}. The unit cells and the orbital degrees of freedom are denoted by $m$ and $A/B$, respectively. The rectangular box represents a particular unit cell. The black dots represent all the connections (or junctions). The expressions (and values) of all the remaining circuit elements are given in the text.}
		\label{circuit3}
	\end{figure}
	
Having explored the TB model and its associated symmetries, let us now shift our focus to constructing a corresponding electric circuit.
The procedure involves a straightforward way of substituting the orbitals of a unit cell of the TB model with nodes or junctions in the topolectrical circuit.
Consequently, the hopping amplitudes or the onsite potentials present in the TB model need to be replaced by specific circuit elements.
For an electrical network, $\mathcal{L}$, consisting of $N$ nodes, if $\mathcal{J}$ denotes the Laplacian \cite{Wu} and $V_i$ and $I_i$ denote the voltage and the current flowing into node `$i$' from the source placed elsewhere respectively, then the following relation must hold,
	\begin{equation}
		I_i=\sum_{j(i\ne i)}X_{ij}(V_i-V_j)+X_iV_i\quad\mathrm{for}\quad j=1, 2, 3.... N,
		\label{eq:current}
	\end{equation}
where $X_{ij}$ is the conductance between two distinct nodes $i$ and $j$.
Note that the term $X_{ii}$ bears no meaning and can be put to zero, whereas $X_i$ is the resultant conductance between node $i$ and the ground.
Thus, Eq.\eqref{eq:current} reads $I=\mathcal{J}V$, where $\mathcal{J}$, $V$ and $I$ denote the Laplacian, the voltage and the current profile of the circuit, respectively.
Consequently, the elements of the Laplacian matrix will be,
    \begin{equation} \mathcal{J}_{ij}=X_{ij}+\delta_{ij}W_i,\quad\mathrm{where}\quad W_i=\sum_{j}X_{ij}+X_i.\label{eq:Lap}
    \end{equation}
Now, one of the measurable quantities for the circuit is the impedance between two nodes, namely $i$ and $j$, which is given by,
	\begin{equation}
		Z_{ij}=\sum_{q_n\ne0}\frac{|\phi_{n,i}-\phi_{n,j}|^2}{q_n},
		\label{im}
	\end{equation}
where $q_n$ is the $n^{\mathrm{th}}$ eigenvalue of the Laplacian $\mathcal{J}$ and $\phi_{n,i}$ is the $i^{\mathrm{th}}$ element of the corresponding eigenmode.

The relationship between the Hamiltonian characterizing the TB model, delineated by Eq.\eqref{eq:Ham1}, and the Laplacian pertaining to the analogous circuit portrayed in Fig.\ref{circuit3}, can be elucidated through the following elaboration.
The hopping parameters $t$ ($-t$) and $t_{AB}$ ($-t_{AB}$) of the TB model are embodied by the capacitors (and inductors) denoted as $C_1$ (and $L_1$) and $C_2$ (and $L_2$), as shown in Fig.\ref{circuit3}.
Furthermore, the modulation of the onsite potential, denoted by $\pm\epsilon$, is deftly achieved by utilising elements $C_3$ and $L_3$.
Note that the circuit elements $C$ (and $L$), which are connected to the ground, are parallel combinations of $C_1$, $C_2$ and $C_3$ (and $L_1$, $L_2$ and $L_3$), respectively, which yields $C=C_1+C_2+C_3$ $\left(\mathrm{and}\;\frac{1}{L}=\frac{1}{L_1}+\frac{1}{L_2}+\frac{1}{L_3}\right)$.
For the present case, we set $|R|=\infty$, which implies that the connection is disjunctive there.
Consider the Laplacian of the Hermitian circuit denoted as $J_{H_1}$.
This Laplacian assumes a specific form using Eq.\eqref{eq:Lap}, and can be expressed as follows:
    \begin{widetext}
    \begin{equation}
    J_{H_1}(\omega) = \begin{pmatrix}
    \frac{1}{i\omega}(\frac{2}{L_1}+\frac{1}{L_2})+ & 0 & \frac{1}{i\omega L_1} & i\omega C_2  & 0 &\dots\dots \\ i\omega (C_2+C)\\
    0 & i\omega(2C_1+C_2)+ & \frac{1}{i\omega L_2} & i\omega C_1  & 0 & \dots\dots \\ & \frac{1}{i\omega}(\frac{2}{L}+\frac{1}{L_2})\\
    \frac{1}{i\omega L_1} & \frac{1}{i\omega L_2} & \frac{1}{i\omega}(\frac{2}{L_1}+\frac{1}{L_2})+ & 0 & \frac{1}{i\omega L_1} & i\omega C_2  & 0 & \dots \\ & & i\omega (C_2+C)\\
    i\omega C_2 & i\omega C_1 & 0 & i\omega(2C_1+C_2)+ & \frac{1}{i\omega L_2} & i\omega C_1  & 0 & \dots \\ & & & \frac{1}{i\omega}(\frac{2}{L}+\frac{1}{L_2})\\
    0 & 0 & \frac{1}{i\omega L_1} & \frac{1}{i\omega L_2} & \frac{1}{i\omega}(\frac{2}{L_1}+\frac{1}{L_2})+ & 0 & \frac{1}{i\omega L_1} & i\omega C_2  & 0 & \dots \\ & & & & i\omega (C_2+C)\\
    \vdots & \vdots & \vdots & \vdots & \vdots & \ddots
    \end{pmatrix}\label{eq:circuit}
    \end{equation}
    \end{widetext}
The Laplacian, $J_{H_1}(\omega_0)$ in Eq.\eqref{eq:circuit}, replicates the Hamiltonian $H_1$ (Eq.\eqref{eq:Ham1}) at a resonance angular frequency $\omega_0$, defined via,
	\begin{equation*}
		\omega_0=2\pi f_0=\frac{1}{\sqrt{L_1C_1}}=\frac{1}{\sqrt{L_2C_2}}=\frac{1}{\sqrt{L_3C_3}}.
	\end{equation*}

\begin{figure}[h!]
		\includegraphics[height=5.5cm, width=9cm]{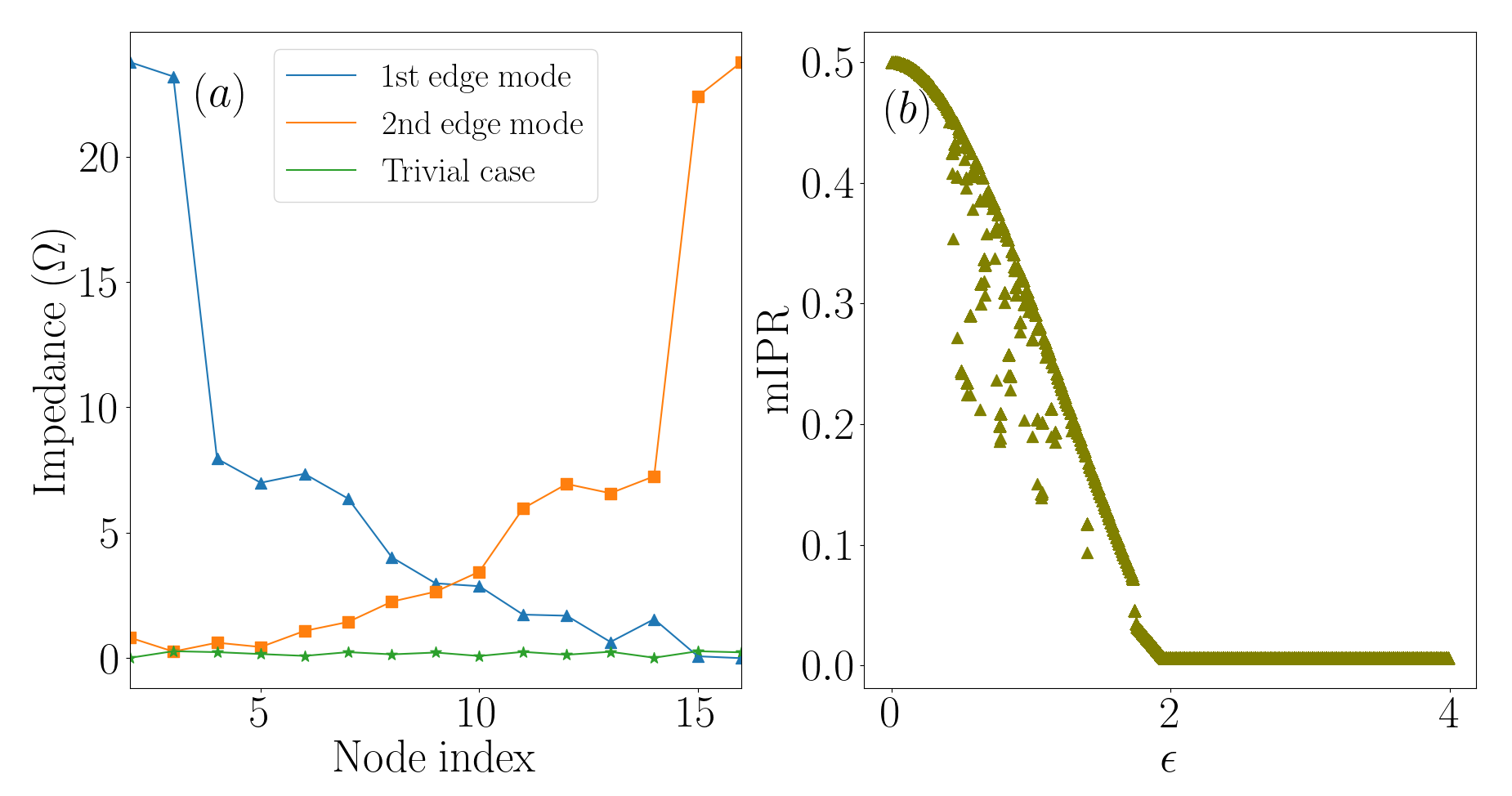}
		\caption{(a) The IP of the Hermitian circuit, comprising eight unit cells, for both the trivial and the topological cases. The first port is fixed at the first(last) node to obtain the second(first) edge mode. We have kept the values of $C_1$, $L_1$, $C_3$ and $L_3$ fixed at $22\;\mu F$, $27\;\mu H$, $6.8\;\mu F$ and $87\;\mu H$, respectively. The resonance frequency of the circuit is $f_0\approx6.53\;kHz$. For the trivial case ($C_2+C_3<3C_1$), we have kept the values of $C_2$ and $L_2$ fixed at $68\;\mu F$ and $8.7\;\mu H$, while for topological case ($C_2+C_3>3C_1$), those have values $33\;\mu F$ and $18\;\mu H$, respectively. (b) mIPR vs $\epsilon$ is shown for the Hermitian TB model comprising $250$ unit cells. The plot suggests that the edge states are non-existent after the point $\epsilon=2t$. Here we have set $t=t_{AB}=1$ and set the energy scale to be in a unit of $t$.} 
		\label{1}
	\end{figure}

Our quest for deeper insights into the electrical circuits and their relevance to topological phenomena leads us to calculate the two-port impedance, in which one port remains anchored at the edge of the circuit while the other is systematically connected to each node within the network.
We obtain impedance measurements at the resonance frequency $\omega_0$ corresponding to each node, as demonstrated in Fig.\ref{1}(a).
Now, the topological characteristics of the TB model will manifest through the emergence of resilient zero-energy boundary modes in real space.
Notably, the IP observed, corresponding to the topological scenario, wherein $C_2+C_3<3C_1$, remarkably mirrors the exponentially decaying probability distribution associated with the zero-energy edge modes.
Consequently, the topological phase transition, that is, an emergence of a trivial phase in our circuit manifests at the critical condition $C_2+C_3=3C_1$.
This corresponds to a spectral gap-closing scenario that occurs for the TB model at $\epsilon=2t$, which shall be discussed later in detail.
Within the trivial regime, the entire profile hovers near zero, indicating that the corresponding eigenmode in the TB model extends uniformly across all the sites.

Now, let us look at the TB model corresponding to OBC.
To establish the localisation characteristics of the eigenstates, we employ a well-known approach involving computation of the inverse participation ratio (IPR) \cite{Kramer} for the TB model, defined via,
	\begin{equation}
		\mathrm{IPR}^{(p)}=\frac{\sum_n|\psi^p_n|^4}{\left(\sum_n|\psi^p_n|^2\right)^2},
		\label{eq:IPR}
	\end{equation}
where $\mathrm{IPR}^{(p)}$ represents the IPR associated with the $p^{\mathrm{th}}$ eigenstate, $\psi^p$, while $n$ signifies the site index.
It is firmly established that for extended states, the IPR exhibits an inverse relationship with the system size ($\sim L^{-1}$), which tends to zero for sufficiently large system sizes.
In contrast, the IPR remains a constant for localised states and is insensitive to the system size.
Further, it approaches unity in the thermodynamic limit where the states are entirely confined to individual sites.
Here, we calculate the `maximum IPR' (mIPR), which denotes the IPR of the edge states in the topological phase and also refers to the highest IPR value among all the eigenstates corresponding to the trivial phase.
The plot of mIPR against $\epsilon$, given in Fig.\ref{1}(b), denotes that the boundary modes exist for $\epsilon<2t$ ($t=1$ in our work) and hence imposes the condition for the topological phase transition to occur at $\epsilon=2t$, which is equivalent to the identity $C_2+C_3=3C_1$ as predicted by the circuit Laplacian.
This equivalence can be understood as the TB potential, $\epsilon$, is expressed via $\epsilon\equiv C_2+C_3-C_1$ and $t\equiv C_1$ as mentioned before.
	
If we consider PBC in the circuit, a Fourier transformation of the Laplacian introduces a wave number component `$k$' per unit cell and spawns a $2\times2$ constitutive block matrix,
    \begin{figure}[h!]
		\centering
		\includegraphics[height=5.5cm, width=9cm]{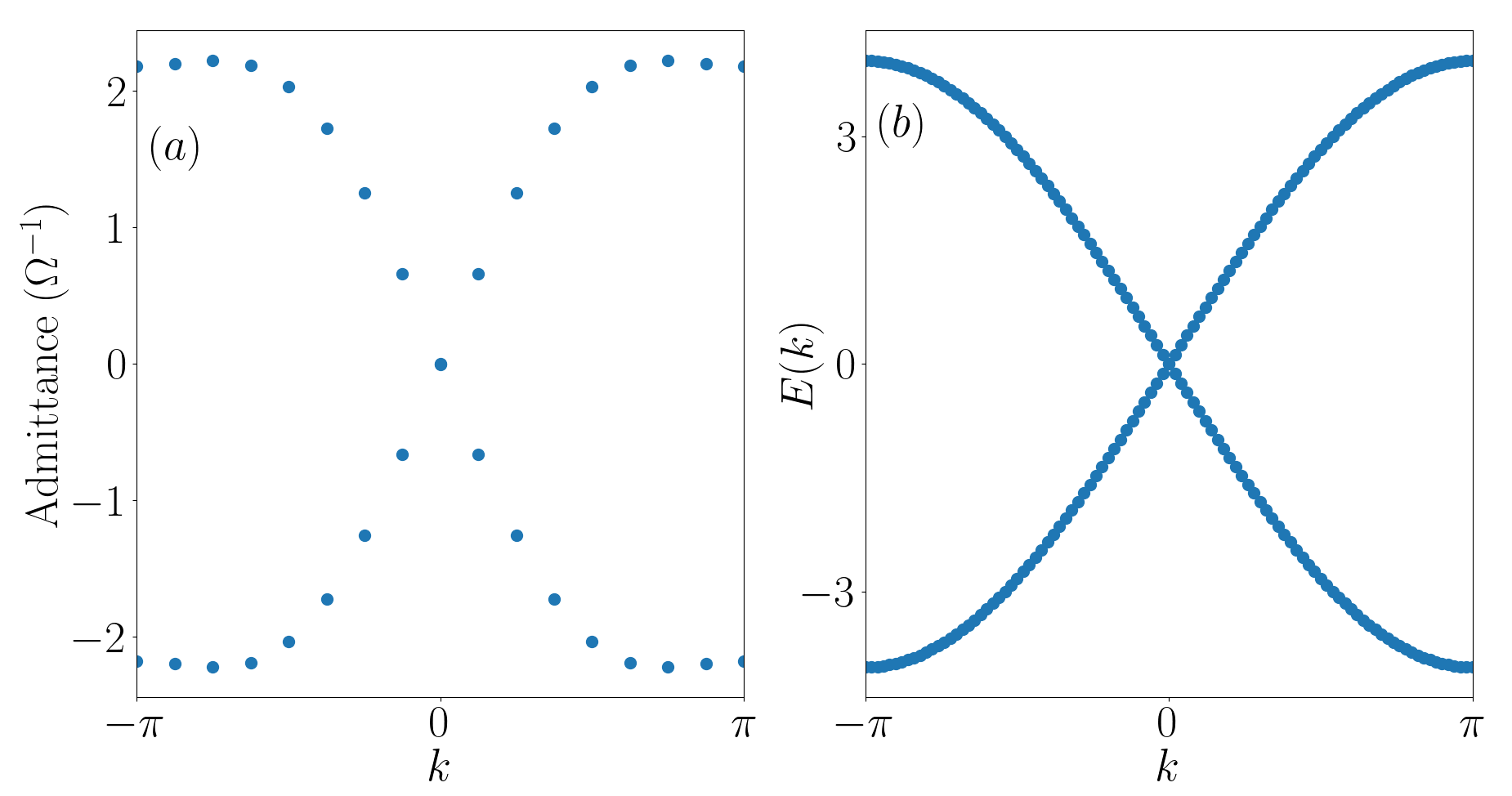}
		\caption{(a) The ABS of the Hermitian ladder circuit with eight unit cells, as described by Eq.\eqref{eq:ckspace_1}. The values of circuit elements are the same as for the circuit with OBC, except for $C_1$ and $L_1$, taking the values $13.3\mu F$ and $44.6\mu H$, respectively. (b) The band structure of the corresponding TB model, comprising of $50$ unit cells, is shown at the phase transition point ($\epsilon=2t$). It correctly replicates the ABS, shown in (a).} 
		\label{2}
	\end{figure}
	\begin{widetext}
	\begin{equation}
		J_{H_1}(\omega,k)=\begin{pmatrix}
			\frac{1}{i\omega L_1}+i\omega(C_2+C_3)+\frac{2}{i\omega L_1}\cos k & -2\omega C_2\sin k\\\frac{2}{\omega L_2}\sin k & i\omega C_1+\frac{1}{i\omega L_2}+\frac{1}{i\omega L_3}+2i\omega C_1\cos k
		\end{pmatrix}.
		\label{eq:ckspace_1}
	\end{equation}
	\end{widetext}
In the absence of any dissipative losses, the spectrum manifests a purely imaginary form; however, in the presence of dissipation, it takes on a complex character.
Upon imposing the PBC on the circuit, the number of certain `significant' nodes \cite{PhysRevB.99.161114} decreases, aligning with the number of sublattices within a unit cell, for example, two in this case.
So, we need only to introduce an input current into two distinct sublattices and measure the voltage responses of the circuit.
Subsequently, we express these responses in terms of Fourier modes, assuming translational invariance of the system.
The combined result from the above scenarios gives rise to an admittance spectrum, which is obtained from the block matrix corresponding to the Laplacian in $k$-space, as presented through Eq.\eqref{eq:ckspace_1}.
The ABS profile is prominently illustrated in Fig.\ref{2}(a), and is symmetric with respect to the value $E(k)=0$, suggesting that the circuit is chirally symmetric.
	
Meanwhile, the band structure of the TB model takes the form $E_{1\pm}(k)=\pm\sqrt{(-\epsilon+2t\cos k)^2+4t_{AB}^2\sin^2k}$ as obtained from Eq.\eqref{eq:kspace_1}, and presented in Fig.\ref{2}(b).
It is imperative to acknowledge that the parameter values employed herein accurately determine the specific location where the topological phase transition occurs.
Consequently, this distinctive choice of parameters reveals a gap closure phenomenon at $k=0$.
Thus, it denotes a critical point in our analysis.
	
\subsection{non-$\mathcal{PT}$-symmetric NH circuit}
	
In this section, we engineer an NH lattice model by introducing a staggered imaginary onsite potential term to the two orbitals.
This modification substitutes $\epsilon$ with $i\epsilon$ in Eq.\eqref{eq:Ham1} for OBC.
This particular form of potential within the Hamiltonian signifies a dynamic exchange of energy, encompassing both `gain' and `loss', a direct consequence of the non-Hermiticity.
In the realm of the $k$-space, the corresponding Bloch Hamiltonian is expressed as follows:
	\begin{equation}
		h_2(k)=\begin{pmatrix}
			i\epsilon-2t\cos k & 2it_{AB}\sin k\\-2it_{AB}\sin k & -i\epsilon+2t\cos k
		\end{pmatrix},
		\label{eq:kspace_2}
	\end{equation}
	which yields the expression for the band structure as,
	\begin{equation}
		E_{2\pm}(k)=\pm\sqrt{(-i\epsilon+2t\cos k)^2+4t_{AB}^2\sin^2k}.
		\label{bs_2}
	\end{equation}
The inclusion of the $i\epsilon$ term disrupts the TRS of the system, that is, $h_2(k)$ does not satisfy Eq.\eqref{eq:TRS}.
In this context, it is noteworthy to mention that the parity operator $\mathcal{P}$ is represented via $\sigma_x$, the $x$-component of the Pauli matrices.
Therefore, in combination with the time-reversal operator $\mathcal{T}$, the $\mathcal{PT}$ operator is expressed as $\mathcal{PT}\equiv\sigma_x\mathcal{K}$.
Thus, $h_2(k)$ is devoid of the $\mathcal{PT}$ or anti-$\mathcal{PT}$ symmetry, mandating that the eigenspectra of the system to assume a complex nature.

The creation of an analogous electric circuit is accomplished by incorporating both positive and negative resistive components ($\pm R$), as illustrated in Fig.\ref{circuit3}, with $R$ being finite for this case, and strategically connected between the circuit nodes and the reference ground.
Furthermore, we have set $C=2C_1$ and $L=\frac{L_1}{2}$ to cancel out any additional terms in the onsite potential other than $\pm R$.
This makes the scenario completely equivalent to $\pm i\epsilon$ in the corresponding TB model.
It is important to note that the implementation of negative impedance is accomplished through negative impedance converters with current inversion, known as INIC \cite{Helbig}.
It works on the principle of the negative feedback configuration of an OP-amp.
A comprehensive discussion and guidelines for using the INICs have been covered in some of the earlier works \cite{PhysRevResearch.4.043108, PhysRevLett.126.215302}.
	\begin{figure}[h!]
		\centering
		\includegraphics[height=5.5cm, width=9cm]{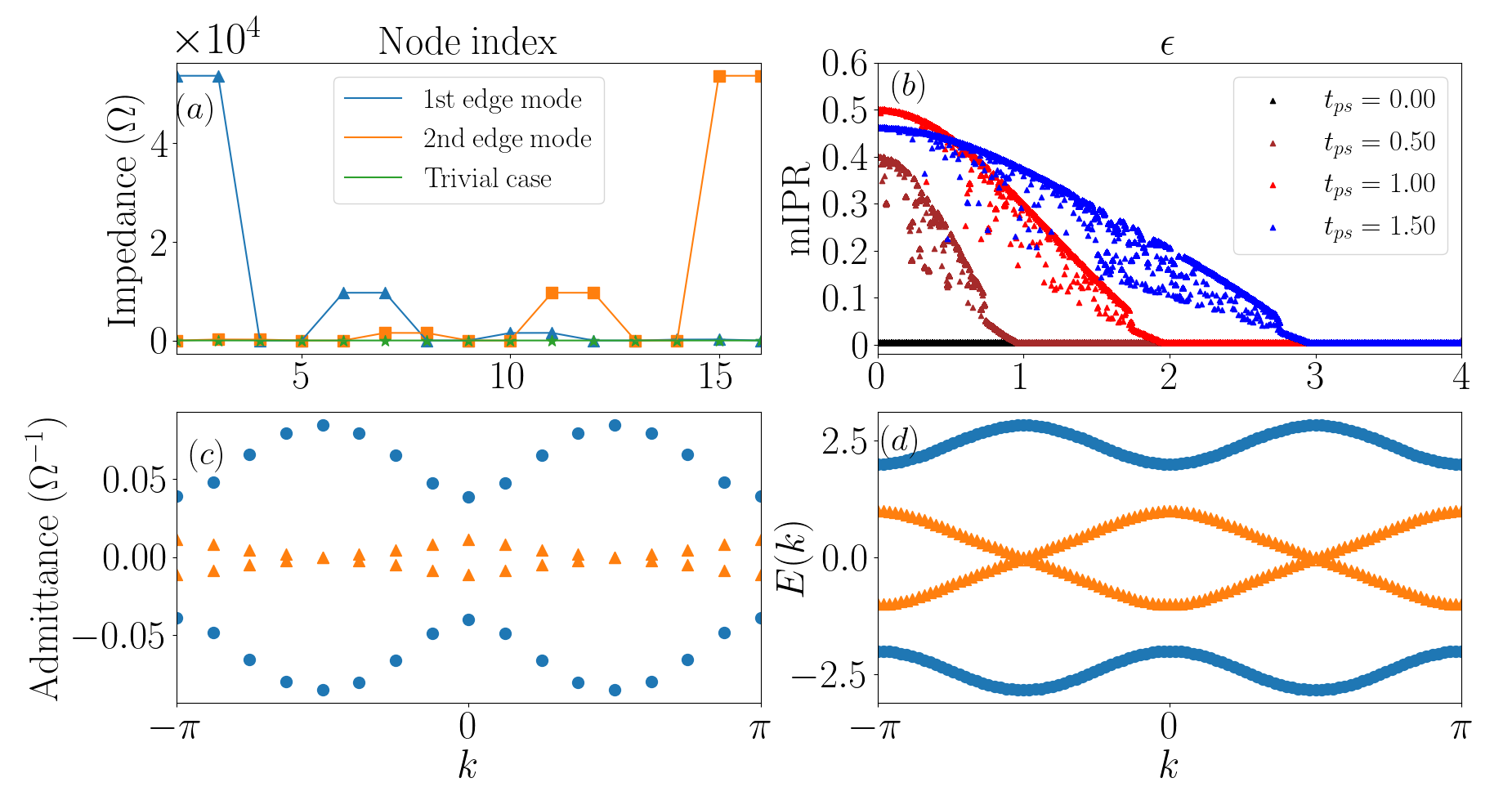}
		\caption{(a) The IP of the non-$\mathcal{PT}$ symmetric NH circuit, constituting of eight unit cells. The trivial and the topological cases correspond to the values of $R$ being $10\;\Omega$ and $100\;\Omega$, respectively. (b) mIPR (defined in the text) versus $\epsilon$ is plotted here for different values of $t_{AB}$ keeping $t=1$. The plot suggests that the existence of the edge states now depends on $t_{AB}$, unlike the Hermitian model, which depends on $t$. The real and imaginary parts of (c) the ABS and (d) the band structure of the corresponding TB model are represented by the circles and triangles, respectively.} 
		\label{3}
	\end{figure}

In Fig.\ref{3}(a), we present the IP for the non-$\mathcal{PT}$-symmetric NH circuit, which has been achieved within both the trivial and topological regions employing the method expounded in the previous section.
To aid the comprehensibility of our results, we have maintained the values of the circuit components identical to that of the Hermitian case, except that the capacitors have been scaled down to the nanofarad ($nF$) range.
Such rescaling allows us a better demonstration of our results on IP and ABS.
Thus, the resonance frequency of the circuit has an approximate value, $f_0 = \frac{\omega_0}{2\pi} \approx 206.5\;kHz$.
The criterion for the topological phase transition in this scenario is succinctly expressed as $R = \frac{1}{2\omega_0C_2} \approx 11.68\;\Omega$.
Fig.\ref{3}(b) showcases the variation of mIPR against $\epsilon$ for four distinct values of $t_{AB}$, namely, $t_{AB} = 0, 0.5, 1\;\mathrm{and}\;1.5$.
These results validate the notion that the phase transition critically hinges upon the interplay between $\epsilon$ and $t_{AB}$, which is analogous to the condition, $R = \frac{1}{2\omega_0C_2}$, for the circuit.
Consequently, mIPR exhibits non-zero values for $\epsilon < 2t_{AB}$, transitioning to zero for $\epsilon > 2t_{AB}$, thereby signifying the topological and trivial regions, respectively.
It is pertinent to mention that the calculation of mIPR follows the same formula as presented in Eq.\eqref{eq:IPR}, albeit with the introduction of the right eigenvectors \cite{PhysRevB.97.045106} tailored to this specific case.
	
For PBC, the Laplacian can be written in Bloch form in the reciprocal space as,
	\begin{equation}
		J_{NH_1}(\omega,k)=\left[\left(\frac{i}{R}+\frac{2}{i\omega L_1}\cos k\right)\sigma_z+\frac{2}{i\omega L_2}\sin k\;\sigma_y\right],
		\label{eq:ckspace_2}
	\end{equation}
where $\sigma_i$ is the $i^{\mathrm{th}}\;(2\times2)$ Pauli matrix.
The inclusion of the resistive element $R$ introduces energy dissipation within the circuit, resulting in a complex admittance profile.
Fig.\ref{3}(c) provides a comprehensive visualisation of the real and imaginary parts of the ABS as a function of the momentum $k$.
Key parameters for this representation include values of $C_1$, $L_1$, and $R$ set at $15\;\mu F$, $39\;\mu H$, and $100\;\Omega$, respectively, while keeping the other parameters consistent with those employed in Fig.\ref{3}(a).
Remarkably, the admittance spectra exhibit a transition into a real domain at the points $k = \pm\frac{\pi}{2}$.
This transformation occurs due to the topological phase that the system resides in.
Consequently, this behaviour faithfully replicates the band structure of the non-$\mathcal{PT}$-symmetric model, as encapsulated by Eq.\eqref{bs_2} and depicted in Fig.\ref{3}(d).
Conducting a spectral analysis of this model reveals the presence of a line gap, a crucial characteristic that substantiates the absence of NHSE \cite{PhysRevLett.124.086801}.
This, in turn, ensures the validation of the bulk-boundary correspondence (BBC) in this particular TB model.
	
\subsection{\label{subs:PT}$\mathcal{PT}$-symmetric NH circuit}

Now, we take recourse to break the Hermiticity of $H_1$ by including a non-reciprocity parameter, $\delta$, in the hopping term among the $A$ and $B$ orbitals ($t_{AB}$) from neighbouring unit cells.
The new Hamiltonian in OBC takes the form,
	\begin{align}
		H_3=H_1+\delta\left[\sum_{i=1}^{L-1}(\hat{a}_{i}^{\dagger}\hat{b}_{i+1}-\hat{b}_{i+1}^{\dagger}\hat{a}_{i})-\sum_{i=2}^{L}(\hat{a}_{i}^{\dagger}\hat{b}_{i-1}-\hat{b}_{i-1}^{\dagger}\hat{a}_{i})\right]
		\label{eq:Ham3}
	\end{align}
with the Bloch Hamiltonian, $h_3(k)$, given by,
	\begin{equation}
		h_3(k)=\begin{pmatrix}
			\epsilon-2t\cos k & 2i(t_{AB}+\delta)\sin k\\-2i(t_{AB}-\delta)\sin k & -\epsilon+2t\cos k
		\end{pmatrix}.
		\label{eq:kspace_3}
	\end{equation}
which indicates that the forward ($A^i\rightarrow B^{i+1}$) and the backward ($A^i\leftarrow B^{i+1}$) hopping amplitudes between $A^i$ and $B^{i+1}$ is $t_{AB}-\delta$ and $t_{AB}+\delta$, respectively, with $i$ being the unit cell index here.
	\begin{figure}[h!]
		\centering
		\includegraphics[height=8cm, width=9cm]{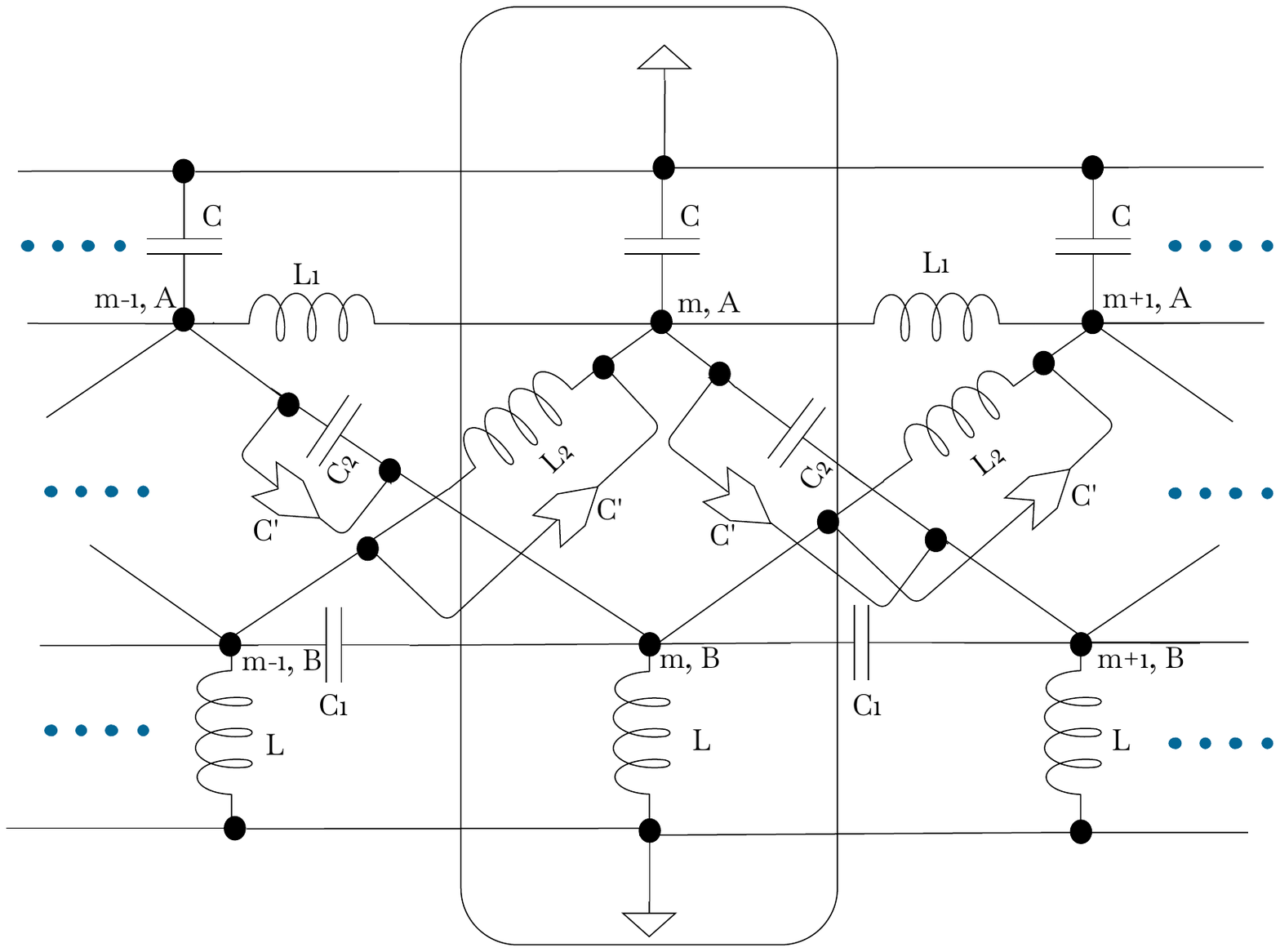}
		\caption{Circuit diagram for $\mathcal{PT}$-symmetric circuit. The INICs are denoted by the arrowheads, which offer an impedance $C'=C_x$ and $C'=-C_x$ for the forward ($m-1,\;A/B\rightarrow m,\;B/A$) and backward ($m,\;A/B\rightarrow m-1,\;B/A$) direction of current, respectively. Thus, the conductance between the nodes becomes $i\omega(C_2\pm C_x)$ from the directional point of view. The expressions (or values) for the remaining circuit elements are given in the text.} 
		\label{circuit2}
	\end{figure}
The non-reciprocity term does not affect the TRS.
The model does not obey $\mathcal{PT}$ symmetry directly, as $h_3(k)$ does not satisfy Eq.\eqref{eq:TRS}.
To establish the $\mathcal{PT}$ symmetry, we shall perform a unitary transformation on $h_3(k)$.
This is achieved via a unitary matrix $U$, such that,
	\begin{gather}
		h_3'(k)=U^{\dagger}h_3(k)U;\quad \mathrm{where}\nonumber\\ U=\frac{1}{\sqrt{2}}
		\begin{pmatrix}
			1 & -1\\1 & 1
		\end{pmatrix},
		\label{eq:U}
	\end{gather}
	which yields,
	\begin{gather}
		h'_3(k)=\begin{pmatrix}
			2i\delta\sin k & -\epsilon+2t\cos k+2it_{AB}\sin k\\-\epsilon+2t\cos k-2it_{AB}\sin k &-2i\delta\sin k
		\end{pmatrix},
		\label{eq:kspace_4}
	\end{gather}
It is now evident that $h'_3(k)$ commutes with the $\mathcal{PT}$ operator in Eq.\eqref{eq:TRS} and is $\mathcal{PT}$-symmetric.
Subsequently, $h_3(k)$ and $h_3'(k)$ are connected via a similarity transformation, and hence the band structure remains identical.
In reference \cite{10.1063/1.1418246}, it was pointed out that a $\mathcal{PT}$-symmetric Hamiltonian can be considered in a broader sense as a specific case of pseudo-Hermiticity, a concept thoroughly discussed in the Appendix \ref{app}.
	\begin{figure}[t!]
		\centering
		\includegraphics[height=5.5cm, width=9cm]{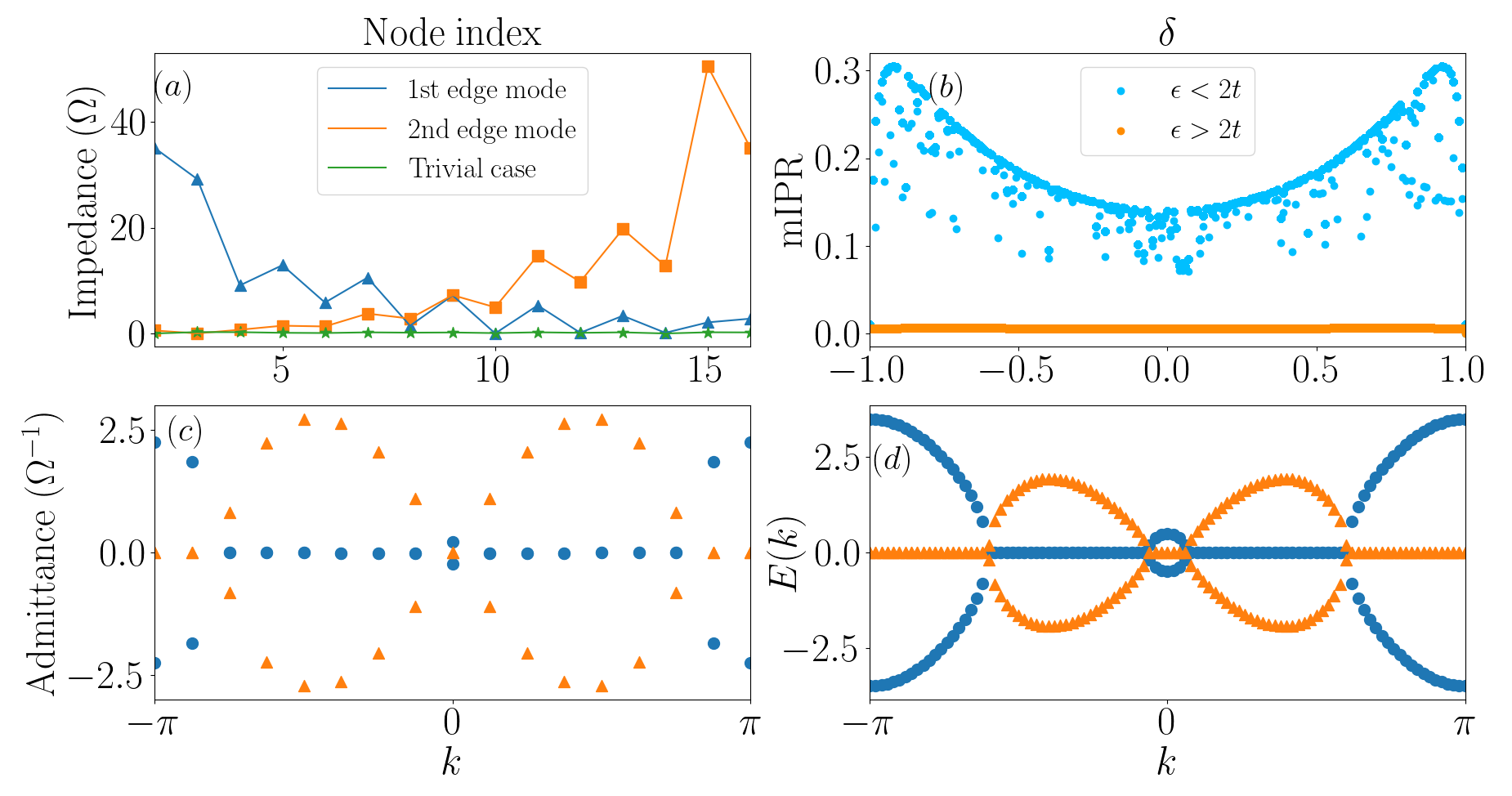}
		\caption{(a) The IP for $\mathcal{PT}$-symmetric NH circuit. The plot is obtained with the same values of circuit elements used for the Hermitian case with OBC and fixing the value of $|C'|$ at $15\;\mu F$, which is always less than $C_2$. (b) The mIPR for $\mathcal{PT}$-symmetric NH lattice model is varied with $\delta$ for $t=t_{AB}=1$ corresponding to the cases $\epsilon=1.5$ ($<2t$) and $\epsilon=3$ ($>2t$), respectively. (c) The ABS of the circuit is shown. All the values of the circuit elements remain the same as the Hermitian circuit, except for $C_1$, $L_1$ and $C',$ which are $15\;\mu F$, $39\;\mu H$ and $34\;\mu F$. (d) The band structure of the lattice model for $\epsilon=\delta=1.5$. The real and imaginary parts of the admittance (and the lattice band structures) are represented by circles and triangles, respectively.} 
		\label{4}
	\end{figure}

The inclusion of non-reciprocity in a circuit is achieved via the INICs, connected in parallel with $C_2$ and $L_2$ between the nodes, as shown in Fig.\ref{circuit2}.
They provide a capacitance $C'=\pm C_x$, equivalent to $\pm\delta$ in the corresponding TB model.
The expressions for $C$ and $L$ remain unchanged from those in the Hermitian circuit.
Surprisingly, the topological phase here depends on two distinct conditions, namely,
	\begin{equation}
		(i)\;C_2+C_3<3C_1;\quad\quad(ii)\;|C'|<C_2,
		\label{con}
	\end{equation}
and violating any of them shall drive the system to a trivial phase.
The IP, mimicking the edge modes, is shown in Fig.\ref{4}(a) and satisfies the above-mentioned conditions (Eq.\eqref{con}).
For our convenience, we have changed the values of $C_2$ and $L_2$, as done for the Hermitian case, to toggle the system between the topological and the trivial phases.
In the corresponding TB model, given by Eq.\eqref{eq:Ham3}, the degree of localisation of these edge states, measured by mIPR, is shown in Fig.\ref{4}(b) as a function of the non-reciprocity parameter, $\delta$, in the range [$-t_{AB}:t_{AB}$] to locate the topological phase transition.
The mIPR is non-zero for $\epsilon<2t$ and supports the existence of the edge states, suggesting that this is a topological phase.
These edge modes vanish as soon as $\epsilon$ becomes larger than $2t$, when all the eigenstates become extended, and mIPR disappears.
These observations validate that the system remains in the topological phase as long as two conditions are met, namely, (i) $\epsilon < 2t$ and (ii) $|\delta| < t_{AB}$.
These conditions are analogous to those specified in Eq.\eqref{con} for the circuit.
In general, the NH systems with non-reciprocity tend to show the presence of NHSE in OBC.
But for this particular system, NHSE is absent.
It is instructive to look at the generic conditions, laid down in Ref.\cite{PhysRevB.99.201103}, for the presence or absence of NHSE in a one-dimensional generalized hopping model.
In addition to this, a detailed explanation for the absence of NHSE is provided in the Appendix \ref{app}, considering the pseudo-Hermitian nature of this model and leveraging insights from the non-Bloch band theory \cite{PhysRevLett.123.066404}.
	
Now, we shall analyse the circuit with PBC through the Bloch Hamiltonian, $J_{NH_2}$, given by,
	\begin{widetext}
		\begin{equation}
			J_{NH_2}(\omega,k)=\left[\left(\frac{1}{i\omega L_1}+i\omega(C_2+C_3)+\frac{2}{i\omega L_1}\cos k\right)\sigma_z+\frac{2}{i\omega L_2}\sin k\;\sigma_y+2i\omega C_x\sin k\;\sigma_x\right].
			\label{eq:ckspace_3}
		\end{equation}
	\end{widetext}
Using this equation, we get the ABS, shown in Fig.\ref{4}(c).
The results clearly show that the admittance values are either purely real or purely imaginary, suggesting the circuit is in $\mathcal{PT}$-broken phase, as the $\mathcal{PT}$-unbroken phase possesses only real eigenvalues.
If we consider the Bloch Hamiltonian of an abstract circuit, denoted as $J_{NH_2}'(\omega, k)$ corresponding to $h_3'(k)$ in Eq.\eqref{eq:kspace_4}, it will yield results similar to those shown in Fig.\ref{4}(c).
This similarity arises because $J_{NH_2}(\omega, k)$ and $J_{NH_2}'(\omega, k)$ are linked through the unitary transformation, which is specified in Eq.\eqref{eq:U} and works for the circuit network as well.
On a parallel front, for the non-reciprocal TB model, the expression for energy, which is the same for both $h_3(k)$ and $h'_3(k)$, is given by,
	\begin{equation}
		E_{3\pm}(k)=\pm\sqrt{(-\epsilon+2t\cos k)^2+4(t_{AB}^2-\delta^2)\sin^2 k}.
		\label{bs_3}
	\end{equation}
The corresponding band structure to the above equation is shown in Fig.\ref{4}(c).
Upon closer examination of Eq.\eqref{bs_3}, we can discern that the condition for the system to exist in the $\mathcal{PT}$-unbroken phase is expressed as $\epsilon > 2\sqrt{\delta^2 + t^2 - t_{AB}^2}$.
This inequality translates to the condition $C_2 + C_3 > C_1 + 2\sqrt{C_x^2 + C_1^2 - C_2^2}$ for the circuit.
The parameters used for both the plots in Figs. \ref{4}(c) and \ref{4}(d) do not satisfy the above conditions, owing to the energy being complex and indicating that they belong to the $\mathcal{PT}$-broken phase.
	
\section{\label{sec3}Conclusion}
	
In this comprehensive study, we have explored topolectrical circuits inspired by a two-orbital, one-dimensional TB model, encompassing both Hermitian and non-Hermitian variants.
The key distinctions between these models lie in their symmetries and the critical values of the parameters, namely, the inter-orbital ($t_{AB}$) and intra-orbital ($t$) hopping amplitudes governing the transition from trivial to topological phases.
While the $\mathcal{PT}$ symmetry is absent in the model with an imaginary onsite potential, it remains intact in the non-reciprocal model.
The localisation characteristics are suitably captured by examining the maximum inverse participation ratio.
Intriguingly, none of these models exhibits any indications of NHSE, despite a natural expectation that non-reciprocal models typically manifest NHSE.
It appears that the non-reciprocal model possesses an inherent pseudo-skew-Hermitian property, leading to the suppression of NHSE and the preservation of the bulk-boundary correspondence in the system.
The observed phenomenon makes it evident that non-reciprocity is not a sufficient condition for the existence of NHSE.
A more in-depth mathematical explanation for this intriguing phenomenon is provided through the non-Bloch band theory.
For the $\mathcal{PT}$-symmetric case, the system hosts purely real eigenvalues in the $\mathcal{PT}$-unbroken regime, while complex eigenvalues emerge in the $\mathcal{PT}$-broken phase.
We have constructed topolectrical circuits for all the three models to bridge the gap between theory and experiment.
The IP of each circuit faithfully mimics the edge modes of the corresponding TB models pertaining to the topological phase.
Additionally, the ABS of the circuit networks with PBC accurately yields the energy band structure of the corresponding models.
In essence, this work offers a perspective on the experimental verification of topological phenomena in NH systems through electronic circuits, shedding light on both the theoretical underpinnings and practical considerations of this intriguing field of study.

\appendix
\section{\label{app}Absence of NHSE in non-reciprocal NH model}
Until now, a substantial body of literature suggests the NHSE is directly connection to the system being non-reciprocal.
Through the reciprocal skin effect found in Ref.\cite{PhysRevResearch.2.023265}, it is known that non-reciprocity is not a necessary condition for NHSE.
What we really illustrate in our manuscript is that non-reciprocity is neither a sufficient condition for NHSE.
In our particular case, the absence of NHSE in a non-reciprocal, yet $\mathcal{PT}$-symmetric, system can be explained through mathematical reasonings.
Let us elaborate on these below.
\subsection{Pseudo-Hermiticity of the Hamiltonian}
NH Hamiltonians with real eigenvalues necessarily belong to a particular class, namely the pseudo-Hermitian (PH) Hamiltonians, which are discussed in details in Ref.\cite{10.1063/1.1418246}.
The condition for pseudo-Hermicity for an arbitrary NH Hamiltonian, $H_{NH}$, can be stated as,
\begin{equation}
    H^{\dagger}_{NH}=\eta\;H_{NH}\;\eta^{-1}
\end{equation}
where $\eta$ is the PH operator and $H_{NH}$ is called $\eta$-PH Hamiltonian.
    \begin{figure}[h!]
        \centering
        \includegraphics[width=0.5\textwidth, height=0.55\columnwidth]{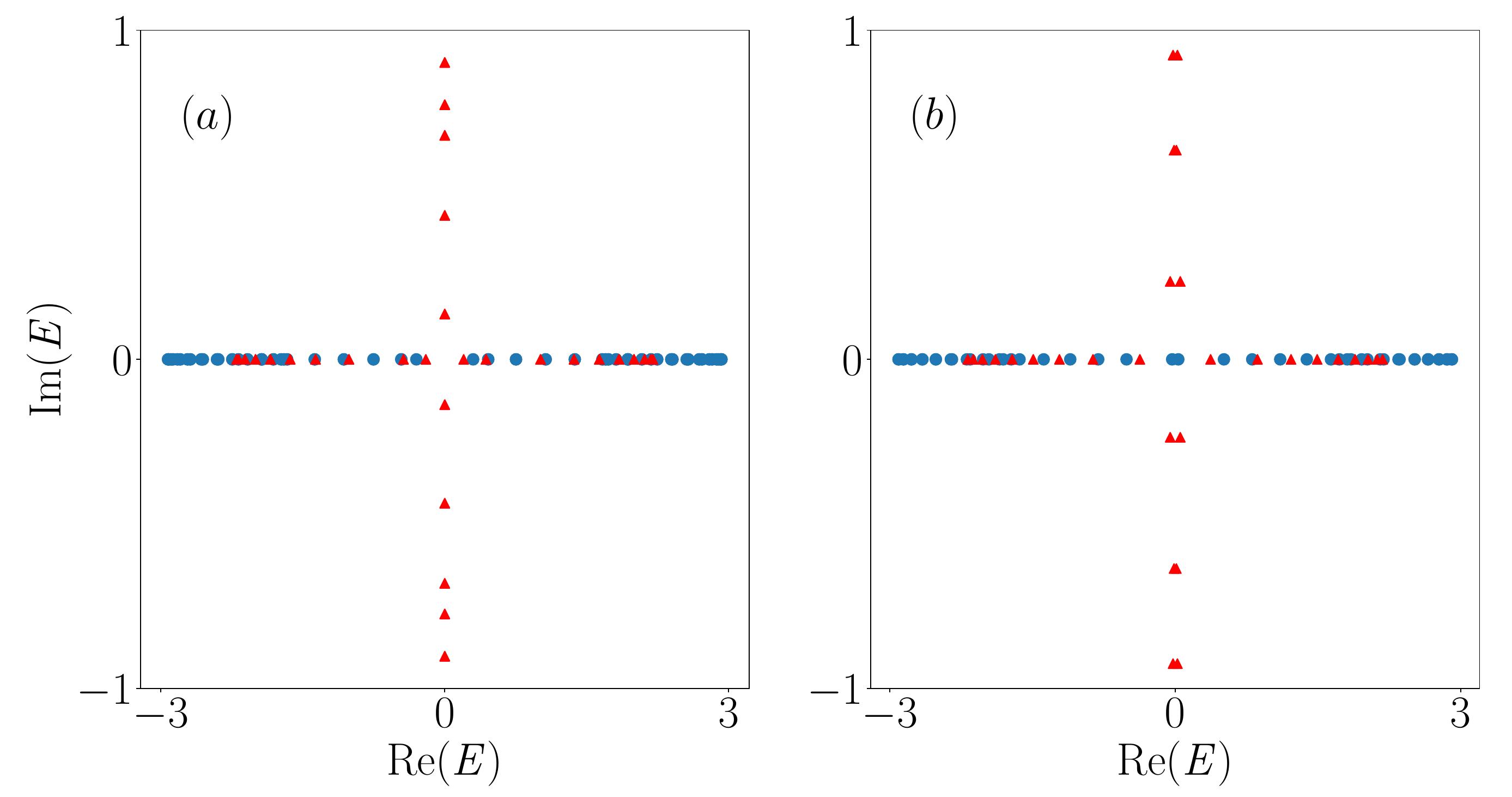}
        \caption{Real and imaginary parts of energy, corresponding to the Hamiltonian $H_3$ (Eq.\eqref{eq:Ham3} of the main text), for PBC (left panel) and OBC (right panel). The blue circles and the red triangles correspond to the cases, $\epsilon<2t$ and $\epsilon>2t$, respectively.}
        \label{rm_n}
    \end{figure}
In our case, the Bloch Hamiltonian for the non-reciprocal model can be written as,
    \begin{equation}
		h_3(k)=\begin{pmatrix}
			\epsilon-2t\cos k & 2i(t_{AB}+\delta)\sin k\\-2i(t_{AB}-\delta)\sin k & -\epsilon+2t\cos k
		\end{pmatrix}.
		\label{eq:kspace_3}
    \end{equation}
Specifically, consider an operator, $\Tilde{\eta}$, for which $h_3(k)$ also satisfies the following relation,
    \begin{equation}
        h^{\dagger}_3(k)=-\Tilde{\eta}\;h_3(k)\Tilde{\eta}^{-1},
        \label{eq:pshh}
    \end{equation}
where $\Tilde{\eta}=U^{\dagger}\sigma_zU$ with $U=\frac{1}{\sqrt{2}}\begin{pmatrix}
        1 & -1\\
        1 & 1
\end{pmatrix}$ and $\sigma_z$ is $z$-component of the Pauli matrix.
Hence, $h_3(k)$ qualifies as a $\Tilde{\eta}$-pseudo-skew-Hermitian ($\Tilde{\eta}$-PSH) Hamiltonian, since it anticommutes with $\Tilde{\eta}$.
By this argument, the system (obeying Eq.\eqref{eq:pshh}) exhibits the behavior of a skew-Hermitian system owing to its pseudo-skew-Hermiticity.
This characteristic effectively suppresses the occurrence of the NHSE in our system.
It is worth mentioning that, we have shown that $h_3(k)$ is $\mathcal{PT}$-symmetric through a unitary transformation via the same operator $U$, that is $h_3'(k)=U^{\dagger}h_3(k)U$.
Since, $h_3'(k)$ respects $\mathcal{PT}$ symmetry, expressed via $\mathcal{PT}\equiv\sigma_x\mathcal{K}$, where $\mathcal{K}$ denotes the complex conjugation operator, the following relation holds,
    \begin{equation}
        h_3'(k)=(\mathcal{PT})h_3'(k)(\mathcal{PT})^{-1}.
    \end{equation}
Hence, Eq. \eqref{eq:pshh} illustrates a more general aspect of the $\mathcal{PT}$ symmetry inherent in $h_3'(k)$.
Adhering to the properties of PH Hamiltonians outlined in Ref.\cite{10.1063/1.1418246}, it can be demonstrated that, for PSH Hamiltonians, one of the following conditions must hold,
    \begin{enumerate}
        \item The eigenvalues of the Hamiltonian are real and come in positive and negative pairs, that is, $(\pm E)$.
        \item The complex eigenvalues come in negative complex conjugate pairs with opposite signs, that is, $(E, -E^*)$.
    \end{enumerate}
Fig.\ref{rm_n} provides a clear visualization of the above conditions.
The blue circles represent the eigenspectra of the Hamiltonian $H_3$ (see Eq.\eqref{eq:Ham3}) in the topological region ($\epsilon<2t$), where the eigenvalues are real and occur in positive and negative pairs.
In contrast, the red triangles are in the trivial region ($\epsilon>2t$), where some eigenvalues are real, occurring in positive and negative pairs, and others are purely imaginary.
The trivial and topological limits are elaborately discussed in Section \ref{subs:PT}.

The essential hallmark of NHSE lies in the sensitivity of the eigenspectra to changes in the boundary conditions of the system \cite{PhysRevB.99.201103}.
The illustration in Fig.\ref{rm_n} clearly depicts that the eigenvalue spectra exhibit insensitivity to boundary conditions (open and periodic), except for the presence of zero energy modes in the topological phase.
This insensitivity is a crucial observation, providing robust evidence that the non-reciprocal Hamiltonian ($H_3$) does not manifest any NHSE.
    
\subsection{Non-Bloch band theory}
The non-Bloch band theory theory dictates the reformulation of the Bloch Hamiltonian $h_3(k)$, in Eq.\eqref{eq:kspace_3}, must be rewritten in terms of $h_3(\beta)$, where $\beta\equiv e^{ik}$ \cite{PhysRevLett.123.066404}.
The ensuing characteristic equation, denoted as $|h_3(\beta)-E\mathbb{1}|=0$ subsequently leads to an equation for a function of $\beta$, namely $f(\beta)$, where 
    \begin{align}
        f(\beta)&=\left(\delta^2+t^2-t_{AB}^2\right)\left(\beta+\frac{1}{\beta}\right)^2-\nonumber\\&2\epsilon t\left(\beta+\frac{1}{\beta}\right)+\epsilon^2+4(t_{AB}^2-\delta^2)=E^2
        \label{eq:beta}
    \end{align}
where $E$ is the energy.
Since this equation is quartic, it admits four solutions for $\beta$, which may be denoted by $\beta_1$, $\beta_2$, $\beta_3$, and $\beta_4$.
To ensure the existence of continuum bands, the following condition must be satisfied,
    \begin{equation}
        |\beta_1|\leq|\beta_2|=|\beta_3|\leq|\beta_4|.
        \label{eq:GBZ}
    \end{equation}
In the framework of Hermitian systems, the generalised Brillouin zone (GBZ), represented as $C_{\beta}$, takes the form of a unit circle in the complex plane.
It is hence crucial to establish that the solutions of Eq.\eqref{eq:beta} fulfill the condition $|\beta_2|=|\beta_3|=1$ to show that it still mimics the behavior of a Hermitian system and hence respects BBC, despite the non-reciprocity of the system.
This assertion implies that the solutions $\beta_2$ and $\beta_3$ lie on a unit circle in the complex plane, akin to the characteristic behavior observed in Hermitian systems.

It will aid if we can decompose $f(\beta)$ in Eq.\eqref{eq:beta} to two quadratic equations.
To this end, let us assume that $\beta$ and $\beta'$ are two distinct solutions (any two out of four in Eq.\eqref{eq:GBZ}), that satisfy Eq.\eqref{eq:beta} and $|\beta|\geq|\beta'|$.
We then assume $\beta=\alpha\beta'e^{i\theta}$, with $\theta$ being real and $\theta\in(0,2\pi)$ and $\alpha\geq 1$.
From Eq.\eqref{eq:beta}, we get $f(\beta)-f(\beta')=0$, which can be written as,
\begin{widetext}
    \begin{equation*}
        \left(\delta^2+t^2-t_{AB}^2\right)\left(\beta+\frac{1}{\beta}\right)^2-2\epsilon t\left(\beta+\frac{1}{\beta}\right)-\left(\delta^2+t^2-t_{AB}^2\right)\left(\beta'+\frac{1}{\beta'}\right)^2+2\epsilon t\left(\beta'+\frac{1}{\beta'}\right)=0
    \end{equation*}
Putting the value of $\beta'$ in terms of $\beta$ in the above equation, we get,
    \begin{equation}
        \left[\frac{\beta e^{-i\theta}}{\alpha}+\frac{\alpha e^{i\theta}}{\beta}-\beta-\frac{1}{\beta}\right]\left[\left(\delta^2+t^2-t_{AB}^2\right)\left(\frac{\beta e^{-i\theta}}{\alpha}+\frac{\alpha e^{i\theta}}{\beta}+\beta+\frac{1}{\beta}\right)-2\epsilon t\right]=0
        \label{eq:b}
    \end{equation}
\end{widetext}
Vanishing of the first square bracket yields,
    \begin{equation}
        \frac{\beta e^{-i\theta}}{\alpha}+\frac{\alpha e^{i\theta}}{\beta}-\beta-\frac{1}{\beta}=0.
        \label{eq:b2}
    \end{equation}
The above equation demands that $\alpha=1$, which confirms the desired conditions namely, $|\beta|=|\beta'|=1$.
Thus, $\beta$ from Eq.\eqref{eq:b2} and $\beta'=\frac{\beta e^{-i\theta}}{\alpha}$ yields $\beta_2$ and $\beta_3$ and confirms that $|\beta_2|=|\beta_3|=1$.
The other solutions namely, $\beta_1$ and $\beta_4$ are obtained for vanishing of the second square bracket in Eq.\eqref{eq:b}.
The mathematical explanation presented above provides a rigorous proof for the Hermiticity of our Hamiltonian, albeit the non-reciprocity, clarifying why the NHSE is not observed in our work.

\end{document}